\documentclass[aps, prx, reprint, showpacs,
               superscriptaddress]{revtex4-1}

\usepackage{mathtools}
\usepackage{siunitx}
\usepackage{dcolumn}

\usepackage{paralist}
\usepackage{verbatim}
\usepackage{afterpage}
\usepackage{siunitx}

\usepackage{microtype}
\usepackage{txfonts}
\usepackage[upright]{txgreeks}
\usepackage{zi4}
\usepackage[mathcal]{euscript}
\usepackage[usenames, dvipsnames, svgnames, table]{xcolor}
\usepackage[colorlinks=true, citecolor=RoyalBlue,
            linkcolor=BrickRed, urlcolor=ForestGreen]{hyperref}

\usepackage{graphicx,calc}
\usepackage[format=plain, justification=centerlast, small]{caption}
\usepackage{subcaption}
\usepackage{tikz}

\newcommand{\rmd}{\mathrm{d}}
\newcommand{\rmdf}{\rmd\hspace{-0.2ex}f}
\newcommand{\rme}{\mathrm{e}}
\newcommand{\rmi}{\mathrm{i}}
\newcommand{\tsup}[1]{^\text{(#1)}}
\newcommand{\bld}{\boldsymbol}
\newcommand{\omag}{\:{\sim}\:}

\DeclareMathOperator{\real}{Re}

\newcommand{\vareps}{\varepsilon}

\newcolumntype{d}[1]{D{.}{.}{#1}}
\newcolumntype{G}{>{\centering\arraybackslash}m{5em}}

\DeclareSIUnit\eV{eV}
\DeclareSIUnit\parsec{pc}
\DeclareSIUnit\yr{yr}

\newcommand{\nm}{\nano\meter}

\newcommand{\km}{\kilo\meter}

\newcommand{\Mpc}{\mega\parsec}

\newcommand{\Hz}{\hertz}
\newcommand{\kHz}{\kilo\hertz}

\newcommand{\rad}{\radian}

\newcommand{\mW}{\milli\watt}

\newcommand{\uN}{\micro\newton}

\begin{document}

\newcommand{\LIGOMIT}{LIGO Laboratory, Massachusetts Institute of Technology, Cambridge, MA 02139, USA}
\newcommand{\LIGOCaltech}{LIGO Laboratory, California Institute of Technology, Pasadena, CA 91125, USA}
\newcommand{\LIGOHanford}{LIGO Hanford Observatory, Richland, WA 99352, USA}
\newcommand{\OzGravMonash}{OzGrav, School of Physics \& Astronomy, Monash University, Clayton 3800, Victoria, Australia}
\title{Systematic calibration error requirements for gravitational-wave detectors via the Cram\'{e}r--Rao bound}
\author{Evan D. Hall}
\email{evanhall@mit.edu}
\affiliation{\LIGOMIT}
\author{Craig Cahillane}
\affiliation{\LIGOCaltech}
\author{Kiwamu Izumi}
\affiliation{\LIGOHanford}
\author{Rory J. E. Smith}
\affiliation{\OzGravMonash}
\author{Rana X Adhikari}
\affiliation{\LIGOCaltech}

\begin{abstract}
Gravitational-wave (GW) laser interferometers such as Advanced LIGO~\cite{Fritschel2014} transduce spacetime strain into optical power fluctuation.
Converting this optical power fluctuations back into an estimated spacetime strain requires a calibration process that accounts for both the interferometer's optomechanical response and the feedback control loop used to control the interferometer test masses.
Systematic errors in the calibration parameters lead to systematic errors in the GW strain estimate, and hence to systematic errors in the astrophysical parameter estimates in a particular GW signal.
In this work we examine this effect for a GW signal similar to GW150914, both for a low-power detector operation similar to the first and second Advanced LIGO observing runs and for a higher-power operation with detuned signal extraction.
We set requirements on the accuracy of the calibration such that the astrophysical parameter estimation is limited by errors introduced by random detector noise, rather than calibration systematics.
We also examine the impact of systematic calibration errors on the possible detection of a massive graviton.
\end{abstract}

\maketitle

\section{Introduction}
\label{sec:Introduction}

Making astrophysical inferences from gravitational-wave detections like GW150914~\cite{GW150914}, GW151226~\cite{GW151226}, and GW170104~\cite{GW170104} requires detector data that is calibrated into spacetime strain with sufficient precision and accuracy~\cite{GW150914Cal}.
Statistical error in the strain calibration has the potential to weaken astrophysical inferences---for example, by increasing the statistical error on the estimated masses of a particular binary system, or weakening constraints on the graviton mass.
Systematic error in the strain calibration, on the other hand, distorts the detector data and therefore has the potential to produce incorrect astrophysical inferences---for example, the graviton mass could be estimated spuriously to be inconsistent with zero.
Constraining systematic calibration errors will become only more pressing with time, as new and improved gravitational-wave detectors see events with ever higher signal-to-noise ratio (SNR), and these events are in turn used to achieve more stringent parameter estimation,.

In certain situations, the effect of systematic calibration errors is straightforward.
In the simple case that one wishes to determine the luminosity distance $D$ of a source using an interferometer whose calibration comprises a simple proportionality constant $k$, then a calibration error $\Delta k/k$ corresponds directly to the error in the estimate of $\Delta D/D$.
However, in the more realistic case that the interferometer calibration function and the GW signal each involve multiple parameters, the relationship between systematic calibration errors and systematic calibration errors is less straightforward.

The effect of calibration errors on GW detection and parameter estimation have focused on placing frequency-dependent constraints on calibration accuracy, without assumptions about the underlying calibration parameters.
Lindblom~\cite{Lindblom2009} derived requirements on the magnitude and phase errors of the interferometer calibration so as to avoid missed signal detections.
Vitale et~al.~\cite{Vitale2012a} examined the effect systematic calibration errors on GW parameter estimation by modeling calibration errors as smooth, random frequency-dependent fluctuations in the interferometer's calibration.

In this work, we first lay out the basic ingredients to Advanced LIGO calibration, including a quasi-zero-pole-gain representation of the optomechanical plant that is valid even in a detuned resonant-sideband-extraction configuration.
We use a semianalytic approach to explicitly relate systematic errors in calibration parameters (e.g., gains, poles and zeros) to systematic errors in GW signal parameters (e.g., masses and distances).
This approach is similar to the approach of Cutler and Vallisneri~\cite{Cutler2007}, who examined how astrophysical parameter estimates are affected by systematic errors in waveform models.

To set requirements on the systematic calibration errors, we compare the systematic calibration-induced errors on the GW signal parameters to the errors induced by the detector's random noise.
Here we use a Fisher-matrix method to estimate the errors due to detector noise via the Cram\'{e}r--Rao bound.
Although this method is known to have limitations~\cite{Vallisneri2008}, its results generally coincide (to within a factor of 2) with Monte Carlo methods for the kind of GW signals expected in Advanced LIGO, so long as the signal-to-noise ratio (SNR) is above ${\sim}20$~\cite{Cokelaer2008,Vitale2012b}.

\section{Interferometer model}
\label{sec:InterferometerModel}

\begin{figure}
    \centering
    \includegraphics[width=\columnwidth]{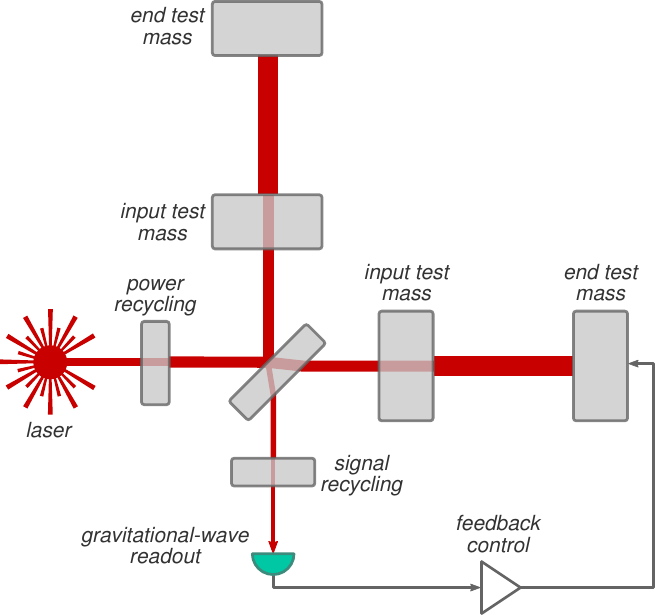}
    \caption{Simplified diagram of Advanced LIGO, showing the optical layout, the GW readout at the antisymmetric port, and the differential arm-length feedback control system.}
    \label{fig:LIGODiagram}
\end{figure}

A basic diagram of an Advanced LIGO interferometer and its differential arm length feedback system is shown in Fig.~\ref{fig:LIGODiagram}.

\subsection{Optomechanical response}
\label{subsec:OptomechanicalResponse}

\begin{figure}[t]
    \centering
    \includegraphics[width=\columnwidth]{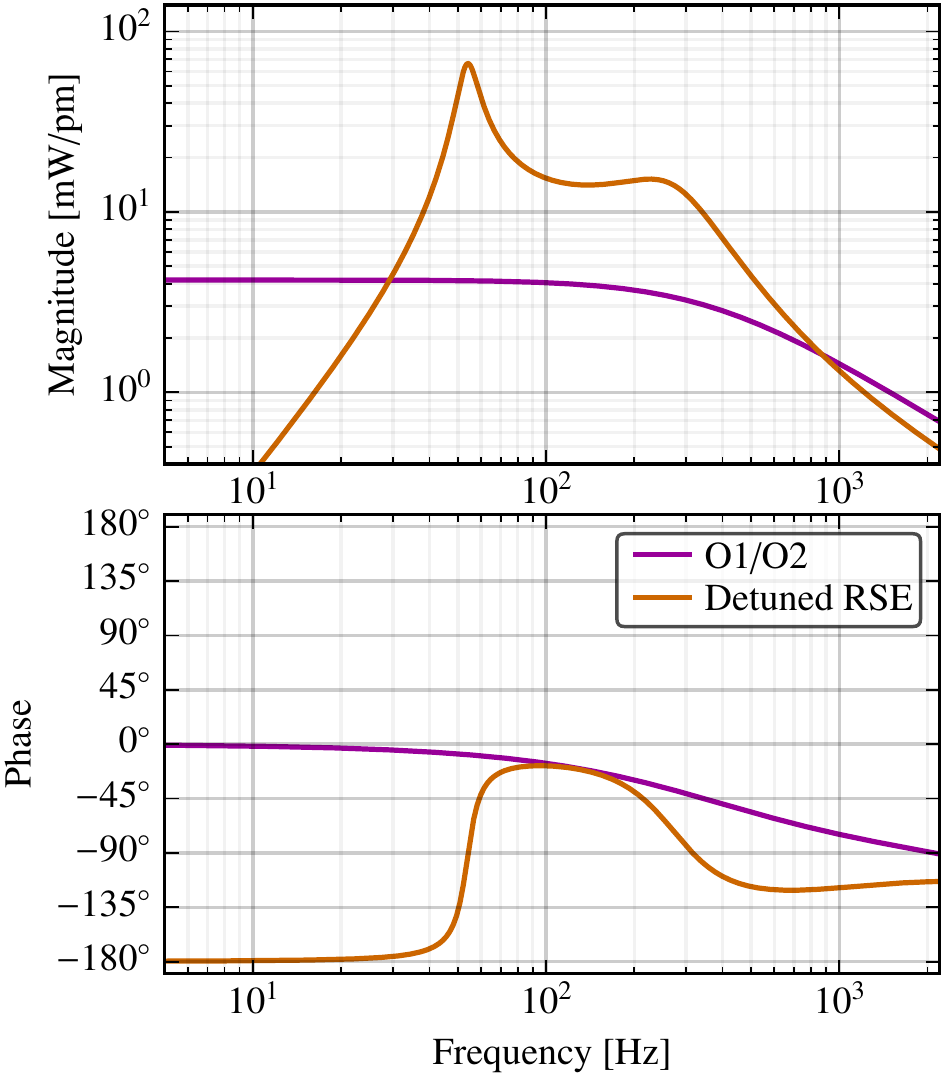}
    \caption{Optomechanical response $C$ of the differential arm length readout for Advanced LIGO, shown for the two configurations considered in this work.}
    \label{fig:OptomechanicalResponses}
\end{figure}

In this section we present the parameters which characterize the optical response of Advanced LIGO (Fig.~\ref{fig:OptomechanicalResponses}).
Each Advanced LIGO detector is a Michelson interferometer whose arms are Fabry--P\'{e}rot cavities; since the end test-masses are negligibly transmissive ($T_\text{e} \omag 5\,\text{ppm}$), the bandwidth of each cavity is set by the input test mass transmissivity $T_\text{i} = 1.4\,\%$, yielding $f_\text{a} = c T_\text{i} / 8\pi L = \SI{42}{Hz}$.
The power circulating in arms is enhanced by the inclusion of a power-recycling mirror (PRM) between the laser and the beamsplitter, and the PRM transmissivity $T_\text{p}$ is chosen to maximize the circulating arm power.

With these six mirrors alone---four test masses, a beamsplitter, and a PRM---the interferometer's optical response in the GW band (between \SI{10}{\Hz} and \SI{7}{\kHz}) would be (to good approximation) a single-pole low-pass filter, with a gain $g$ set by the arm power and the optical losses, and a single pole $p$ equal to the arm bandwidth $f_\text{a}$.
However, each Advanced LIGO detector additionally employs a scheme called resonant sideband extraction (RSE), in which a signal recycling mirror (SRM) is placed at the detector's antisymmetric port to alter the detector's optical response~\cite{Mizuno1993} and its quantum-limited noise performance~\cite{Buonanno2001}.
The exact nature of the alteration depends on the SRM's power transmissivity $T_\text{s}$ and the microscopic signal-recycling-cavity (SRC) length $\ell_\text{s}$.
Additionally, the interferometer optical response is affected by the homodyne angle $\zeta$, which describes the audio-band demodulation quadrature of the GW readout.

In the end, the interferometer optical response is determined by the following five physical parameters:
\begin{enumerate}
    \item the power $P = \vareps\sqrt{P_\text{bs} P_\text{lo}}$, which depends on the circulating power $P_\text{bs}$ impinging on the beamsplitter, the local oscillator power $P_\text{lo}$ used to detect the GW signal, and any optical losses $\varepsilon$;
    \item the arm bandwidth $f_\text{a}$;
    \item the power transmissivity $T_\text{s}$ of the SRM;
    \item the microscopic one-way SRC phase $\phi = (2\pi\ell_\text{s}/\lambda_0) \mod{2\pi}$; and
    \item the homodyne angle $\zeta$.
\end{enumerate}

However, the optical response is characterized more directly via a quasi-zero-pole-gain representation comprising the following parameters.
\begin{enumerate}
    \item The optical gain $g$ (with units of watts per meter), which depends on the beamsplitter power $P_\text{bs}$, the local oscillator power $P_\text{lo}$, and any optical losses in the system.
    \item The homodyne zero
        \begin{equation}
            z = f_\text{a} \times \frac{\cos(\phi+\zeta) - r_\text{s} \cos(\phi-\zeta)}{\cos(\phi+\zeta) + r_\text{s} \cos(\phi-\zeta)}.
        \end{equation}
    \item[\stepcounter{enumi}\theenumi--\stepcounter{enumi}\theenumi.] The pole
        \begin{equation}
            p = f_\text{a} \times \frac{1 - r_\text{s}\rme^{2\rmi\phi}}{1 + r_\text{s}\rme^{2\rmi\phi}},
        \end{equation}
        which (being complex) comprises two independent parameters.
        We will find it most convenient to work with the magnitude
        \begin{equation}
            |p| = f_\text{a}\times\left(\frac{1-2r_\text{s}\cos2\phi+r_\text{s}^2}{1+2r_\text{s}\cos2\phi+r_\text{s}^2}\right)^{1/2}
        \end{equation}
        and the $Q$ factor
        \begin{equation}
            Q_p = \frac{|p|}{2\real{p}} = \frac{1}{2} \times \frac{\left(1-2r_\text{s}^2\cos4\phi + r_\text{s}^4\right)^{1/2}}{1-r_\text{s}^2},
        \end{equation}
        which attains a minimum value of 1/2 when $p$ is real.
    \item The squared spring frequency
        \begin{equation}
            \xi^2 = f_\text{a}^2 \times \frac{2r_\text{s}\sin2\phi}{1-2r_\text{s}\cos2\phi+r_\text{s}^2} \times \frac{c P_\text{bs}/\lambda_0}{2\pi^3 f_\text{a}^4 M L^2},
        \end{equation}
        where $\lambda_0 = \SI{1064}{\nm}$ is the laser wavelength, and $M = \SI{40}{\kg}$ is the mass of the interferometer test masses.
        $\xi^2$ is positive in the presence of optical spring and negative in the presence of optical antispring.
\end{enumerate}
With these five parameters---$g$, $z$, $|p|$, $Q_p$, and $\xi^2$---the interferometer's optomechanical response $C(f)$ is
\begin{equation}
    C(f) = \frac{g\,\rme^{-2\pi\rmi f L/c}\times(1+\rmi f/z)}{1 + \rmi f/|p|Q_p - f^2/|p|^2 - \xi^2/f^2},
    \label{eq:OptomechanicalPlant}
\end{equation}
where the factor $\rme^{-2\pi\rmi f L/c}$ accounts for the time delay of signals propagating down the arms.

In this work we consider two different realizations of the optomechanical response for Advanced LIGO.
These are shown in Fig.~\ref{fig:OptomechanicalResponses}, and the corresponding parameters are given in Tab.~\ref{tab:Configurations}.

The first configuration corresponds to the Advanced LIGO detectors as they were operated during the first and second observing runs (O1 and O2).
Here the detectors are operated with extremal RSE ($\phi = \SI{90}{\degree}$)~\footnote{In the Hanford detector, a slight optical antispring is observed, with $\xi^2 \simeq -(\SI{7}{\Hz})^2$ or $\phi \simeq \SI{90.5}{\degree}$, but this is not considered for the simulations in this work.}, with \SI{25}{\watt} of input power, an SRM with transmissivity $T_\text{s} = \SI{37}{\%}$, and a homodyne readout angle $\zeta = \SI{90}{\degree}$.
In this configuration, the optical response $C(f)$ reduces to
\begin{equation}
    C\tsup{O1/O2}(f) = \frac{g\,\rme^{-2\pi\rmi f L/c}}{1+\rmi f/|p|},
    \label{eq:OptomechanicalPlantESE}
\end{equation}
and hence requires characterization of only $g$ and $|p|$, since the remaining parameters $z$, $Q_p$, and $\xi^2$ are then known.

The second configuration corresponds to a possible future observing run in which the detectors may be operated with detuned RSE to optimize the signal-to-noise ratio for compact binary coalescence signals.
Here the detectors are operated with a one-way SRC phase $\phi = \SI{82}{\degree}$, with \SI{125}{\watt} of input power, $T_\text{s} = \SI{20}{\%}$, and a homodyne angle $\zeta = \SI{100}{\degree}$.
The optomechanical plant in this configuration is given by the full expression in Eq.~\ref{eq:OptomechanicalPlant}, and hence requires characterization of $g$, $z$, $|p|$, $Q_p$, and $\xi^2$.

\begin{table}[t]
    \centering
    \caption{Detector calibration parameters and astrophysical signal parameters for the simulations performed in this paper.
        Note that for the zero-detuned O1/O2 configuration, the quantities $z$, $Q_p$ and $\xi^2$ are not included as separate parameters in the optical response model: $z$ is assumed to be equal to $|p|$, and $Q_p$ and $\xi^2$ are assumed to be known exactly.}
    \begin{ruledtabular}
    \begin{tabular}{c d{1.1} d{1.1} l}
            \textbf{Quantity}  &
            \multicolumn{1}{c}{\textbf{O1/O2}}    &
            \multicolumn{1}{c}{\textbf{Detuned RSE}}  &
            \textbf{Unit} \\
        \hline
        $g$     & 4.2    &   9.8   & mW/pm    \\
        $z$     & 365    & 749     & Hz       \\
        $|p|$   & 365    & 276     & Hz       \\
        $Q_p$   & 0.50   &   1.33  & ---      \\
        $\xi^2$ & 0.0    & +53.4^2 & Hz$^2$   \\
        $a$     & 0.11   & 0.11    & $\mu$N/V \\
        \hline
        $\mathcal{M}$    & \multicolumn{2}{d{1.1}}{30.7}  & $M_\odot$ \\
        $\eta$           & \multicolumn{2}{d{1.1}}{0.247} & --- \\
        $t_\text{c}$     & \multicolumn{2}{d{1.1}}{2.50}  & ms \\
        $\phi_\text{c}$  & \multicolumn{2}{d{1.1}}{0.600} & rad \\
        $D$              & \multicolumn{2}{d{1.1}}{756}   & Mpc \\
        $m_G$     & \multicolumn{2}{d{1.1}}{0}     & kg \\
        \hline
        $\rho$  & 23     & 63      & ---      \\
    \end{tabular}
    \end{ruledtabular}
    \label{tab:Configurations}
\end{table}

\subsection{Feedback control loop}
\label{subsec:FeedbackControl}

\begin{figure}[t]
    \centering
    \includegraphics[width=\columnwidth]{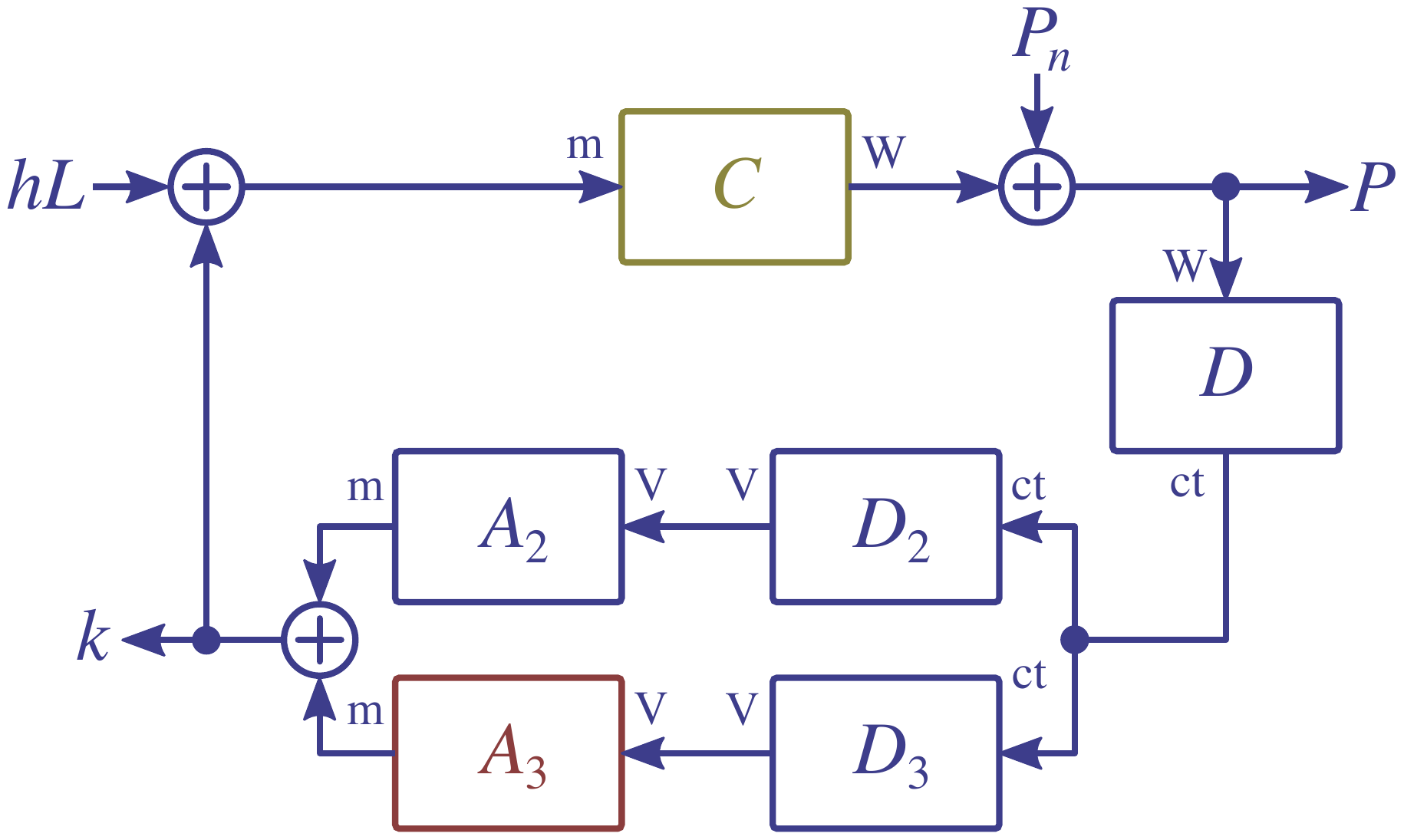}
    \caption{Loop diagram of the differential arm length feedback control.
        The interferometer's optomechanical response $C$ converts astrophysically-induced displacement $hL$ into power fluctuation $P_h$, which is summed with the detector's noise $P_n$ to produce the total observed power fluctuation $P$.
        This fluctuation is digitized and filtered by the functions $D$, $D_2$, and $D_3$ to produce a set of control voltages for the multi-stage test mass suspensions.
        These voltages are applied to the electromechanical length actuators for each suspension stage ($A_2$ and $A_3$), producing a displacement control signal $k$ which acts to suppress differential arm length fluctuations in the interferometer.
        }
    \label{fig:DarmLoopDiagram}
\end{figure}

The optomechanical response of the interferometer is not the only transfer function that must be accounted for to produce an estimated strain signal.
Because the interferometer's differential arm length is actively servoed by feeding the GW readout back to the test masses, the effect of this servo loop must be accounted for.
A diagram of the servo loop is shown in Fig.~\ref{fig:DarmLoopDiagram}.

We consider the loop in three parts.
The first is the optomechanical response $C(f)$ (already described in Sec.~\ref{subsec:OptomechanicalResponse}) which converts differential arm length displacement $L_-(f)$ into power fluctuation $P(f)$ at the interferometer's dark port.
The second is a set of electronic transfer functions $\{D, D_1(f), D_2(f), D_3(f)\}$, which take the power fluctuation $P(f)$ and produce a set $\{k_1(f), k_2(f), k_3(f)\}$ of control voltages intended to be fed back to three of the test mass suspension actuators.
The third is a set $\{A_1(f), A_2(f), A_3(f)\}$ of actuator transfer functions which describe how each control voltage produces displacement of the test mass.
The total open-loop transfer function $G(f)$ of the interferometer's differential arm length servo is
\begin{equation}
    G(f) = CD\times(D_2A_2 + D_3 A_3),
    \label{eq:DarmOLTF}
\end{equation}
where we have ignored the first-stage suspension actuation term $D_1 A_1$, since its main effect is to suppress length fluctuations at frequencies below ${\sim}\SI{1}{\Hz}$, which is below the GW band.

In this paper we assume that all of the electronic, digital, and mechanical parameters that characterize $D$, $D_2$, $D_3$, $A_2$, and $A_3$ are fixed and known to negligible uncertainty \emph{except} the actuation strength $a$ (in newtons per volt) that determines the overall magnitude of the bottom-stage test mass transfer function $A_3$.
The bottom-stage actuator is an electrostatic drive, and is affected by the accumulation of free charges on or near the test mass~\cite{Martynov2016}.
The distribution of charge near the test mass is known to vary with time.
Therefore, the actuation strength $a$ is included along with $g$, $z$, $|p|$, $Q_p$, and $\xi^2$ as a detector calibration parameter.
In this work, its magnitude is fixed at $a = \SI{0.11}{\uN/\volt}$ for both the O1/O2 and detuned-RSE configurations.

\begin{figure}[t]
    \centering
    \includegraphics[width=\columnwidth]{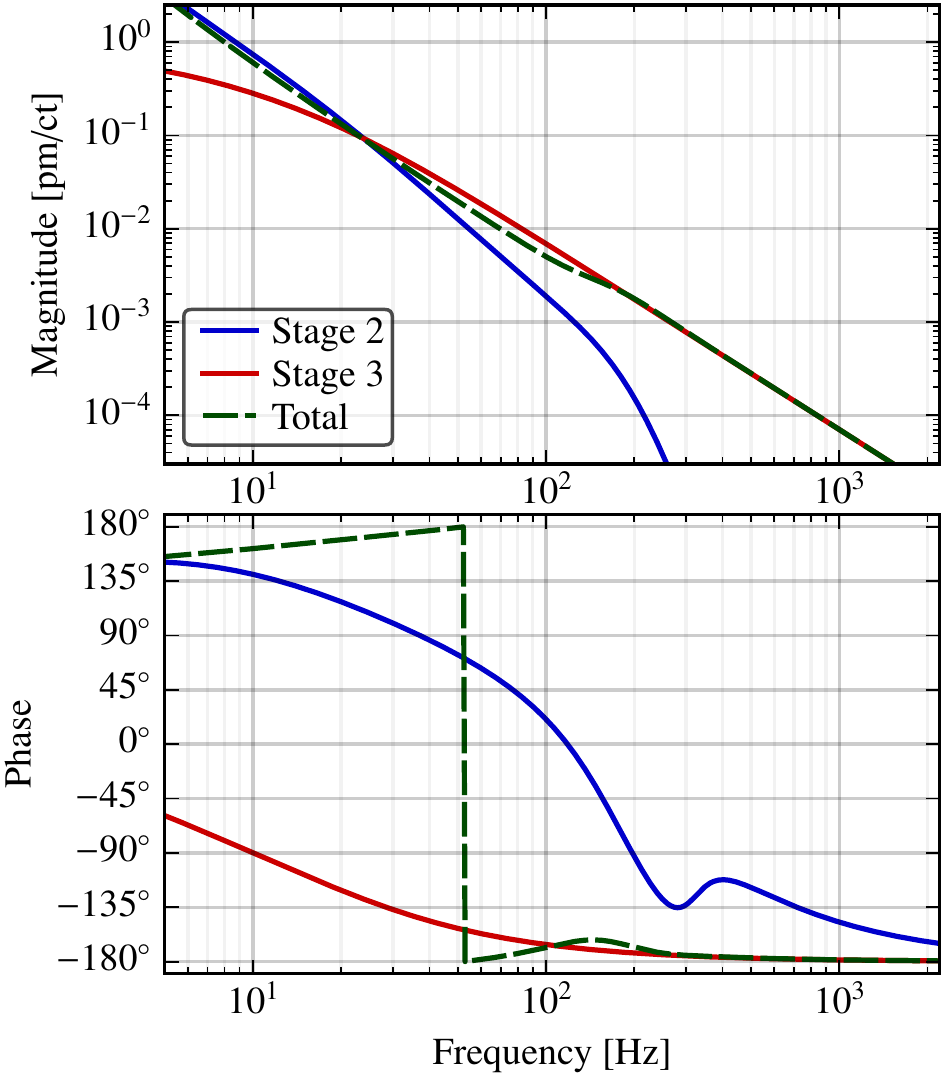}
    \caption{Actuation transfer function $D_2 A_2 + D_3 A_3$ for the differential arm length loop (the low-frequency first stage $D_1 A_1$ has been omitted).
    The overall strength of stage 3 is known to drift because of the charge distribution in the vicinity of the test mass; therefore, this strength $a$ (in newtons per volt) must be carefully measured as part of the calibration process.}
    \label{fig:DarmActuation}
\end{figure}

\begin{figure}[t]
    \centering
    \includegraphics[width=\columnwidth]{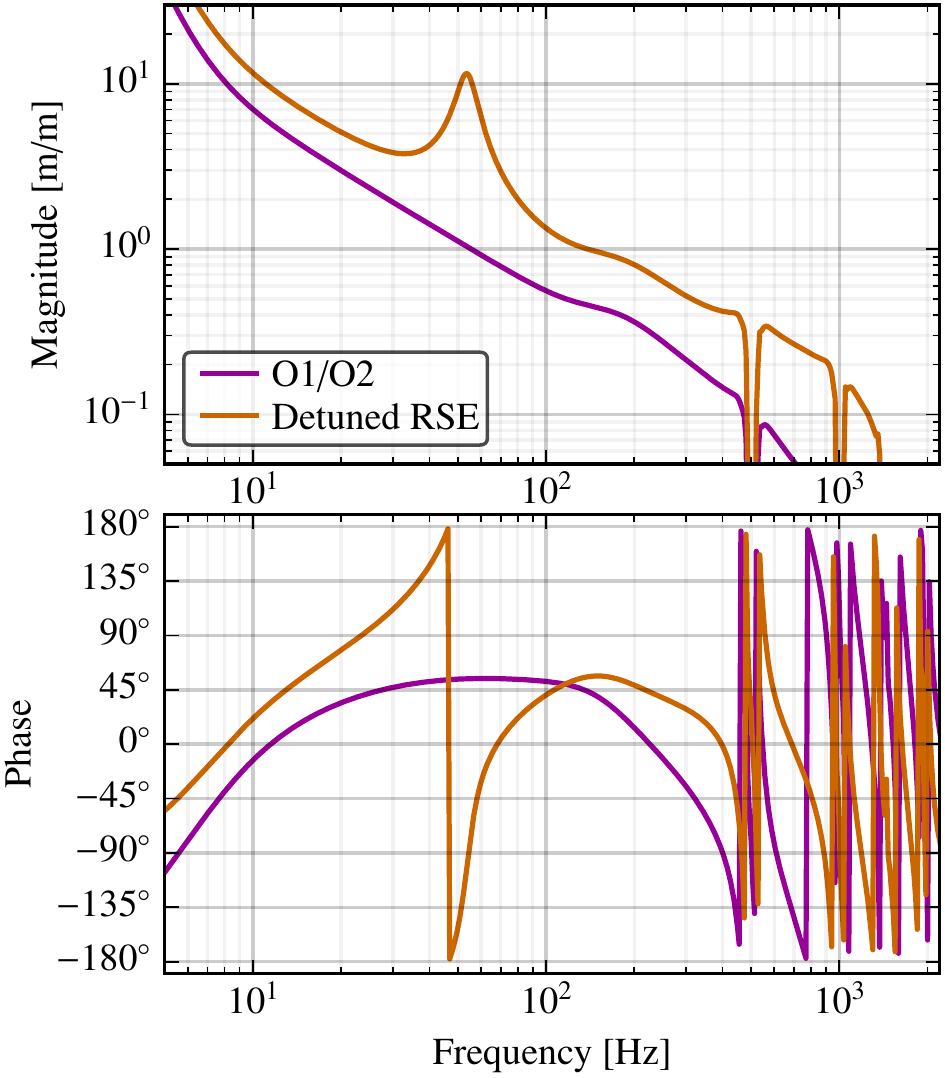}
    \caption{Open-loop transfer function $G$ of the differential arm length loop for the two configurations considered here (Eq.~\ref{eq:DarmOLTF}).}
    \label{fig:DarmOLTF}
\end{figure}

\section{Strain estimation}
\label{sec:StrainEstimation}

The estimated freerunning strain $d(f)$ is produced from the measured power fluctuation $P(f)$ via
\begin{equation}
    d(f) = \frac{1}{L}\times \frac{1-G(f)}{C(f)} \times P(f),
    \label{eq:d(f)}
\end{equation}
where $C(f)$ and $G(f)$ have been described in Sec.~\ref{sec:InterferometerModel}, and $L = \SI{3995}{\meter}$ is the average arm length.
In the GW literature, the quantity $[1-G(f)]/C(f)$ is frequently called the ``response function'' and is denoted $R(f)$~\footnote{In this work, we use the convention that the servo goes unstable if $G(f) = +1$.
If instead the opposite convention is used [instability occurs for $G(f) = -1$], then the response function $R(f)$ has the form $[1+G(f)]/C(f)$.}.

We rewrite Eq.~\ref{eq:d(f)} as
\begin{equation}
    d(f) = \frac{R(f)}{L} \times [P_h(f) + P_n(f)],
    \label{eq:StrainFromPower}
\end{equation}
where we have split $P(f)$ into two components: $P_h(f)$, which accounts for power fluctuation induced by astrophysical strain, and $P_n(f)$, which accounts for power fluctuation induced by detector noise.
In this equation, it is apparent that the estimated strain $d(f)$ is not equal to the \emph{true} strain $h(f)$ incident on the detector, for two reasons.
First, the presence of $P_n$ introduces random noise in the data on top of the true astrophysical fluctuation $P_h$.
Second, the estimated response function $R(f)$ may differ from the true response function $R_\text{t}(f)$ because of systematic calibration errors.
Errors in the estimate of $R$ will therefore cause errors in the estimated freerunning strain $d$.

Similarly, the estimated strain noise power spectral density (PSD) $S_{\! nn}(f)$ is related to the detector's power noise PSD $S_{\! P_n P_n}(f)$ by
\begin{equation}
    S_{\! nn}(f) = \frac{|R(f)|^2}{L^2} \times S_{\! P_n P_n}(f),
    \label{eq:PsdFromPower}
\end{equation}
and hence is similarly susceptible to systematic errors in $R(f)$.

\section{Parameter estimation}
\label{sec:ParameterEstimation}

We now examine how systematic errors in the interferometer calibration affect the estimation of astrophysical parameters from a compact binary coalescence.

\subsection{General formalism}

We suppose we have some frequency-domain data $d(f)$, collected from one detector only, that is known to contain a coalescence signal.
On the other hand, we have a model that produces a frequency-domain waveform $h(f;\bld{\theta})$; here $\bld{\theta}$ is a vector of parameters (component masses, coalescence time, etc.) describing the astrophysical event.
The goal of parameter estimation is to find the parameters $\bld{\hat{\theta}}$ that best match the detector data $d$.
In this work we will focus on choosing $\bld{\hat{\theta}}$ via maximum-likelihood (ML) estimation, although the procedure described below could be extended to Bayesian estimation~\cite{vonToussaint2011}.
ML estimation requires the construction of a likelihood function $\mathcal{L}$, which is proportional to $p(d|\bld{\theta})$, the probability of observing the detector data $d$ given that the astrophysical event has parameters $\bld{\theta}$.

\begin{figure}[t]
    \centering
    \includegraphics[width=\columnwidth]{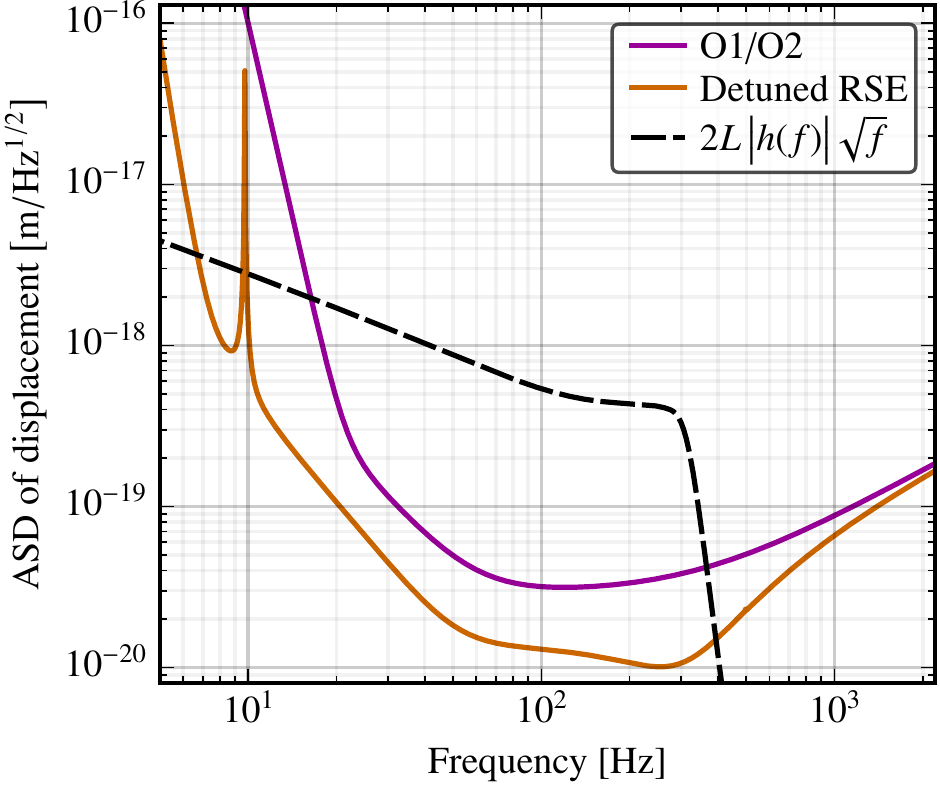}
    \caption{Displacement-referred amplitude spectral density (ASD) $\sqrt{S_{\! nn}(f)}\times L$ for differential arm length noise, both for an O1/O2 configuration and for a possible future detuned RSE configuration.
    To facilitate comparison with the strength of the GW signal $h(f)$ described in Sec.~\ref{sec:Results}, we also plot the quantity $2L\,\bigl|h(f)\bigr|\sqrt{f}$.}
    \label{fig:LIGONoise}
\end{figure}

If the detector's noise is stationary and Gaussian, the logarithm of $\mathcal{L}$ is given by
\begin{equation}
    \ell \equiv \ln\mathcal{L}
         = -\frac{1}{2}\int\limits_0^\infty \!\! \rmdf \; \frac{\bigl|d(f) - h(f;\bld{\theta})\bigr|^2}{S_{\! nn}(f)},
\end{equation}
where $d(f)$ is the frequency-domain detector data, $h(f;\bld{\theta})$ is the frequency-domain waveform model (which is a function of the parameters $\bld{\theta}$), and $S_{\! nn}(f)$ is the power spectral density (PSD) of the detector's noise.
ML estimation proceeds by maximizing the log-likelihood $\ell$ with respect to $\bld{\theta}$, yielding the ML estimate $\bld{\hat{\theta}}$.

Several quantities we consider later will depend on defining the usual noise-weighted, frequency-domain inner product between two sets $s_1(f)$ and $s_2(f)$ of data:~\cite{Creighton2011}
\begin{equation}
    \langle s_1 | s_2 \rangle = 4 \int\limits_0^\infty\!\! \rmdf\; \frac{\real{\bigl[s_1^*(f) \,s_2(f)\bigr]}}{S_{\! nn}(f)}.
    \label{eq:InnerProduct}
\end{equation}
In particular, the best matched-filter SNR that could be obtained for a waveform $h(f)$ is
\begin{equation}
    \rho = \sqrt{\langle h | h \rangle}.
    \label{eq:SNR}
\end{equation}
Additionally, the Fisher matrix $\bld\Gamma$ for a waveform $h(f)$ which depends on parameters $\bld{\theta}$ has elements~\cite{Creighton2011}
\begin{equation}
    \Gamma_{ij} = \left\langle \frac{\partial h}{\partial\theta_i} \middle| \frac{\partial h}{\partial\theta_j} \right\rangle.
    \label{eq:FisherMatrix}
\end{equation}
The Fisher matrix describes the extent to which detector noise introduces random errors in the estimate of $\bld{\theta}$.
The Cram\'{e}r--Rao bound implies that the covariance matrix $\bld\Sigma$ of these errors is bounded elementwise by the inverse of $\bld\Gamma$:~\cite{Creighton2011}
\begin{equation}
    \Sigma_{ij} \ge \bigl(\bld{\Gamma}^{-1}\bigr)_{ij}.
    \label{eq:CramerRao}
\end{equation}

\subsection{Effect of calibration errors}

Both $d(f)$ and $S_{\! nn}(f)$ are also implicitly functions of the calibration parameters, which we will denote $\bld{\lambda}$.
Because of systematic errors, the calibration parameters $\bld{\lambda}$ which are used to produce the strain data from power fluctuations at the GW readout port will \emph{not} necessarily correspond to the detector's true calibration parameters $\bld{\lambda}_\text{t}$; rather, they will differ by some amount $\Delta\bld{\lambda} = \bld{\lambda} - \bld{\lambda}_\text{t}$.
Using Eqs.~\ref{eq:StrainFromPower} and~\ref{eq:PsdFromPower}, we can write the estimated strain data and estimated strain PSD as
\begin{equation}
    d(f;\bld{\lambda}) = \frac{R(f;\bld{\lambda})}{R(f;\bld{\lambda}_\text{t})} \times d(f;\bld{\lambda}_\text{t})
\end{equation}
and
\begin{equation}
    S_{\! nn}(f;\bld{\lambda}) = \left|\frac{R(f;\bld{\lambda})}{R(f;\bld{\lambda}_\text{t})}\right|^2 \times S_{\! nn}(f;\bld{\lambda}_\text{t}),
\end{equation}
where $d(f;\bld{\lambda}_\text{t})$ and $S_{\! nn}(f;\bld{\lambda}_\text{t})$ are the estimated strain data and estimated strain PSD that would have been produced in the absence of systematic calibration errors.

With these equations we can rewrite the log-likelihood explicitly as a function of both $\bld{\theta}$ and $\bld{\lambda}$:
\begin{equation}
    \ell = -\frac{1}{2}\int\limits_0^\infty \!\! \frac{\rmdf}{S_{\! nn}(f;\bld{\lambda}_\text{t})} \times \left| d(f;\bld{\lambda}_\text{t}) - \frac{R(f;\bld{\lambda}_\text{t})}{R(f;\bld{\lambda})} h(f;\bld{\theta})\right|^2\!\!.
\end{equation}
Viewed in this way, we see that if $\bld{\lambda}$ is shifted from its true value $\bld{\lambda}_\text{t}$ by an amount $\Delta\bld{\lambda}$, then the log-likelihood $\ell$ will shift, and hence the ML estimate $\bld{\hat{\theta}}$ will shift from its true value $\bld{\hat{\theta}}_\text{t}$ by an amount $\Delta\bld{\hat{\theta}}$.

We now compute how $\Delta\bld{\hat{\theta}}$ depends on $\Delta\bld{\lambda}$.
For this calculation, we assume that the signal-to-noise ratio of the GW signal is strong, so that $|P_h(f)| \gg |P_n(f)|$, and that the waveform model $h(f;\bld{\theta})$ used to match the detector data $d(f;\bld{\lambda})$ is free of any modeling errors or unmodeled parameters.
This ensures that in the absence of calibration errors, we have $d(f;\bld{\lambda}) = d(f;\bld{\lambda}_\text{t}) = h(f;\bld{\theta}_\text{t})$, where $\bld{\theta}_\text{t}$ are the system's true parameters.
The ML estimate will then be $\bld{\hat{\theta}} = \bld{\theta}_\text{t}$, and hence $\ell$ attains its maximum value of $0$.

\begin{widetext}
We now examine how the ML estimate $\Delta\bld{\hat{\theta}}$ shifts in the presence of nonzero systematic calibration errors $\Delta\bld{\lambda}$.
For definiteness we say that $\bld{\theta}$ consists of $M$ parameters, and $\bld{\lambda}$ consists of $N$ parameters.
First, we allow both $\bld{\lambda}$ and $\bld{\theta}$ to vary freely, so that to second order the total change in the log-likelihood is
\begin{equation}
    \Delta\ell = \sum_i \frac{\partial\ell}{\partial\mu_i} \Delta\mu_i + \frac{1}{2}\sum_i\sum_j \frac{\partial^2\ell}{\partial\mu_i\partial\mu_j} \Delta\mu_i\,\Delta\mu_j,
\end{equation}
where $\mu_i,\mu_j \in \{\theta_1,\theta_2,\ldots,\theta_M,\lambda_1,\lambda_2,\ldots,\lambda_N\}$.
More explicitly,
\newcommand{\evalopt}{{\bld{\theta}_\text{t},\bld{\lambda}_\text{t}}}
\begin{equation}
    \Delta\ell = \sum_i \left.\frac{\partial\ell}{\partial\lambda_i}\right|_\evalopt{}
                    \hspace{-1ex} \Delta\lambda_i \hspace{1ex}
               + \sum_i \left. \frac{\partial\ell}{\partial\theta_i}\right|_\evalopt{}
                    \hspace{-1ex} \Delta\theta_i \hspace{1ex}
               + \frac{1}{2}\sum_i\sum_j \left.\frac{\partial^2\ell}{\partial\lambda_i\partial\lambda_j}\right|_\evalopt{}                                \hspace{-1ex} \Delta\lambda_i\,\Delta\lambda_j \hspace{1ex}
               + \frac{1}{2}\sum_i\sum_j \left.\frac{\partial^2\ell}{\partial\theta_i\partial\theta_j}\right|_\evalopt{}
                    \hspace{-1ex} \Delta\theta_i\,\Delta\theta_j \hspace{1ex}
               + \sum_i\sum_j \left.\frac{\partial^2\ell}{\partial\lambda_i\partial\theta_j}\right|_\evalopt{}
                    \hspace{-1ex} \Delta\lambda_i\,\Delta\theta_j.
    \label{eq:DelLogLike}
\end{equation}

We now fix the systematic calibration errors $\Delta\lambda_1, \Delta\lambda_2, \ldots, \Delta\lambda_N$.
Since we noted earlier that $\ell$ attains its maximum value of $0$ when $\Delta\bld{\lambda} = 0$, it must be the case that $\Delta\ell$ is nonpositive, and hence the shift $\Delta\bld{\hat{\theta}}$ will maximize $\Delta\ell$.
The particular values $\Delta\bld{\hat{\theta}}$ will satisfy the $M$ equations
\begin{equation}
\hspace{7em}    0 = \frac{\partial\Delta\ell}{\partial\Delta\theta_k}
      = \left.\frac{\partial\ell}{\partial\theta_k}\right|_\evalopt{}
        + \sum_j \left.\frac{\partial^2\ell}{\partial\theta_k\partial\theta_j}\right|_\evalopt{}
            \hspace{-1ex} \Delta\theta_j \hspace{1ex}
        + \sum_i \left.\frac{\partial^2\ell}{\partial\theta_k\partial\lambda_i}\right|_\evalopt{}
            \hspace{-1ex} \Delta\lambda_i, \hspace{4em} k \in \{1, 2, \ldots, M\}.
\end{equation}
We assumed earlier that $\ell$ attains its maximum for $\bld{\theta} = \bld{\theta}_\text{t}$ and $\bld{\lambda} = \bld{\lambda}_\text{t}$.
Therefore, we must have $\partial\ell/\partial\theta_k\big|_\evalopt{} = 0$.
Therefore, the relation between $\Delta\bld{\hat{\theta}}$ and $\Delta\bld{\lambda}$ is
\begin{equation}
   -\underbrace{
    \begin{bmatrix}
        \ell_{\theta_1\theta_1} & \ell_{\theta_1\theta_2} & \cdots & \ell_{\theta_1\theta_M} \\
        \ell_{\theta_2\theta_1} & \ell_{\theta_2\theta_2} & \cdots & \ell_{\theta_2\theta_M} \\
        \vdots & \vdots & \ddots & \vdots \\
        \ell_{\theta_M\theta_1} & \ell_{\theta_M\theta_2} & \cdots & \ell_{\theta_M\theta_M}
    \end{bmatrix}}_{\equiv\bld{\mathcal{H}}}
    \begin{bmatrix}
        \Delta\hat{\theta}_1 \\ \Delta\hat{\theta}_2 \\ \vdots \\ \Delta\hat{\theta}_M
    \end{bmatrix}
    =
    \underbrace{
    \begin{bmatrix}
        \ell_{\theta_1\lambda_1} & \ell_{\theta_1\lambda_2} & \cdots & \ell_{\theta_1\lambda_N} \\
        \ell_{\theta_2\lambda_1} & \ell_{\theta_2\lambda_2} & \cdots & \ell_{\theta_2\lambda_N} \\
        \vdots & \vdots & \ddots & \vdots \\
        \ell_{\theta_M\lambda_1} & \ell_{\theta_M\lambda_2} & \cdots & \ell_{\theta_M\lambda_N}
    \end{bmatrix}}_{\equiv\bld{\mathcal{M}}}
    \begin{bmatrix}
        \Delta\lambda_1 \\ \Delta\lambda_2 \\ \vdots \\ \Delta\lambda_N
    \end{bmatrix},
    \label{eq:CeMatrices}
\end{equation}
with
\begin{equation}
    \ell_{\mu_i \mu_j} = \left.\frac{\partial^2\ell}{\partial\mu_i\partial\mu_i}\right|_\evalopt{}
\end{equation}
and $\mu_i,\mu_j \in \{\theta_1,\theta_2,\ldots,\theta_M,\lambda_1,\lambda_2,\ldots,\lambda_N\}$.
\end{widetext}

In Eq.~\ref{eq:CeMatrices} we note that the matrix $\bld{\mathcal{H}}$ on the left-hand side is the Hessian of $\ell$ with respect to $\bld{\theta}$.
Additionally, to the matrix on the right-hand side we have assigned the letter $\bld{\mathcal{M}}$.
This allows us to write the relationship between calibration errors $\Delta\bld{\lambda}$ and parameter estimation errors $\Delta\bld{\hat{\theta}}$ as
\begin{equation}
    \Delta\bld{\hat{\theta}} = \mathbf{J}\,\Delta\bld{\lambda}
\end{equation}
with
\begin{equation}
    \mathbf{J} = -\bld{\mathcal{H}}^{-1}\bld{\mathcal{M}}.
    \label{eq:CeJac}
\end{equation}
As desired, the matrix $\mathbf{J}$ in Eq.~\ref{eq:CeJac} quantifies how systematic calibration errors $\Delta\bld{\lambda}$ produce a shift $\Delta\bld{\hat{\theta}}$ in the ML estimate of the astrophysical signal parameters.
The rest of this work will be dedicated to computing $\mathbf{J}$ for a GW150914-like coalescence signal.

\subsection{Choice of waveform}

We use a family of phenomenological waveforms~\cite{Husa2016,Khan2016} that incorporates the dynamics of the inspiral, merger, and ringdown phases of the coalescence.
This family also includes effects from the spins of the components, but in this work we constrain the spins to be zero.
The five parameters we consider are
\begin{enumerate}
    \item the chirp mass $\mathcal{M} = (m_1 m_2)^{3/5} \big/ (m_1+m_2)^{1/5}$, where $m_1$ and $m_2$ are the component masses;
    \item the symmetric mass ratio $\eta = m_1 m_2 \big/ (m_1 + m_2)^2$;
    \item the coalescence time $t_\text{c}$;
    \item the coalescence phase $\phi_\text{c}$; and
    \item the effective distance $D$, which depends on both the actual system distance and the orientation of the source relative to the detector.
\end{enumerate}
We additionally assume that the source is oriented so that the detector senses the strain from only the plus-polarization of the GW.

In parts of the next section, we will additionally consider the problem of GW parameter estimation under the assumption of a massive graviton.
The effect of a massive graviton on a post-Newtonian waveform was considered by Will~\cite{Will1998}.
The energy $E$, momentum $p$, and mass $m_G$ of the graviton are related by the dispersion relation $E^2 = p^2 c^2 + m_G^2 c^4$, and from this it can be shown that the GW waveform $h(f)$ acquires an extra phase term
\begin{equation}
    \psi_G(f) = -\frac{\mathcal{D} m_G^2 c^3}{4\pi\hbar^2(1+Z)f} = - \frac{\pi\mathcal{D}c}{\Lambda^2(1+Z)f} \equiv -\frac{B}{f},
    \label{eq:MgPhase}
\end{equation}
where $\Lambda = 2\pi\hbar/m_G c$ is the Compton wavelength of the graviton,  $\mathcal{D}$ is a cosmological distance quantity (not equal in general to the luminosity distance $D$), $Z$ is the source redshift, and we have defined $B = \pi\mathcal{D}c/\Lambda^2(1+Z)$.

\section{Results}
\label{sec:Results}

We consider a coalescence signal with parameters $\mathcal{M} = 30.7\,M_\odot$, $\eta = 0.247$, $t_\text{c} = \SI{2.50}{\ms}$, $\phi_\text{c} = \SI{0.600}{\rad}$, $D = \SI{756}{\Mpc}$.
The values of $\mathcal{M}$ and $\eta$ are similar to the binary-black-hole signal GW150914~\cite{GW150914,O1BBH}; the value of $D$ is chosen to give an overall strain amplitude similar to that of GW150914; and the values of $t_\text{c}$ and $\phi_\text{c}$ are chosen arbitrarily.
Additionally, we assume the system is at a redshift $Z = 0.09$ (also similar to GW150914) and that the graviton is massless ($m_G = B = 0$; $\Lambda = \infty$).

\subsection{O1/O2 configuration}

Here we consider a coalescence detection when the detector is operating in its extremal RSE configuration, for which the optomechanical plant is given by Eq.~\ref{eq:OptomechanicalPlantESE}.
In this case, the vector of calibration parameters is $\bld{\lambda} = \begin{pmatrix} g & |p| & a \end{pmatrix}^\intercal$, where $g$ and $|p|$ are the gain and pole of the optomechanical plant, and $a$ is the actuation strength of the test mass actuator.
The true calibration parameters are assumed to be $g = \SI{4.2}{\mW/pm}$, $|p| = \SI{365}{\Hz}$, and $a = \SI{0.11}{\uN/\volt}$ (Table~\ref{tab:Configurations}).
In this configuration, the signal has a matched-filter SNR $\rho = 23$.

In the rest of this section we calculate the effect of systematic calibration errors on the parameter estimation of this signal in two cases: first, in the case that our parameter estimation assumes a massless graviton; and second, in the case that our parameter estimation includes the graviton mass as a parameter to be estimated from the signal.

\subsubsection{Massless graviton}

Here we consider parameter estimation assuming a massless graviton ($\Lambda = \infty$ and $B = 0$), so that the vector of astrophysical parameters is $\bld{\theta} = \begin{pmatrix} \mathcal{M} & \eta & t_\text{c} & \phi_\text{c} & D \end{pmatrix}^\intercal$.

Applying Eq.~\ref{eq:CeJac} to this scenario yields the following relationship between $\Delta\bld{\hat\theta}$ and $\Delta\bld{\lambda}$:
\begin{equation}
    \begin{bmatrix}
        \Delta \mathcal{M}/\mathcal{M} \\
        \Delta \eta /\eta \\
        \Delta t_\text{c} / (\SI{1}{\ms}) \\
        \Delta \phi_\text{c} / (\SI{1}{\rad}) \\
        \Delta D/D
    \end{bmatrix}
    =
    \frac{1}{10^3}\times
    \left[
    \begin{array}{d{5.1} d{5.1} d{5.0}}
        -12 & -21 & 1 \\
        -93 & -80 & -181 \\
        -1745 & -121 & -2827 \\
        -5058 & -10497 & 7770 \\
        736 & -90 & -537 \\
    \end{array}
    \right]
    \begin{bmatrix}
        \Delta g/g \\
        \Delta |p|/|p| \\
        \Delta a/a
    \end{bmatrix},
    \label{eq:CeJacZeroDetNoMG}
\end{equation}
where we have expressed the relationships fractionally where appropriate.

We want to use the result in Eq.~\ref{eq:CeJacZeroDetNoMG} to set reasonable goals on the systematic errors in $g$, $|p|$, and $a$.
As noted by Lindblom~\cite{Lindblom2009}, once the calibration-induced systematic errors $\Delta{\bld{\hat{\theta}}}$ are made sufficiently small, the parameter estimation will be dominated by systematics induced by detector noise, and more stringent calibration will not help.
In this work, we will therefore set the calibration requirements so that the $\Delta{\bld{\hat{\theta}}}$ is no more than one third of these noise-induced systematic errors.

To estimate the typical size of noise-induced systematic errors, we compute the Cram\'{e}r--Rao bound on the covariance matrix of $\bld{\hat{\theta}}$ via Eqs.~\ref{eq:FisherMatrix} and~\ref{eq:CramerRao}.
In this case, the bound implies that the diagonal elements of the covariance matrix can be no smaller than the following values:
\begin{subequations}
\begin{align}
		\Sigma_{\mathcal{M}\mathcal{M}}^{1/2}/\mathcal{M} &= 0.015 \\
		\Sigma_{\eta\eta}^{1/2}/\eta &= 0.051 \\
		\Sigma_{t_\text{c} t_\text{c}}^{1/2} &= 0.45\,\text{ms} \\
		\Sigma_{\phi_\text{c} \phi_\text{c}}^{1/2} &= 8.3\,\text{rad} \\
		\Sigma_{DD}^{1/2}/D &= 0.054.
\end{align}
\end{subequations}
We then use these values to set a requirement on the systematic calibration errors $\Delta\bld{\lambda}$ by requiring that the calibration errors introduce a systematic error $\Delta\bld{\hat{\theta}}$ that is less than one third that the Cram\'{e}r--Rao limit.
The results of this requirement are given in Tab.~\ref{tab:CalRequirements}.

\begin{table}[t]
    \centering
    \caption{Requirements on systematic calibration errors for a GW150914-like signal.
    These requirements are set such that the systematic calibration errors induce an error in the astrophysical parameter estimation that is less than one third of the error introduced by detector noise (as determined from the Cram\'{e}r--Rao bound as described in the text).}
    \begin{ruledtabular}
    \begin{tabular}{c d{2.2} d{2.2} d{2.2} d{2.2}}
            \textbf{Quantity}
            & \multicolumn{1}{G}{\textbf{O1/O2, no MG}}
            & \multicolumn{1}{G}{\textbf{O1/O2, MG}}
            & \multicolumn{1}{G}{\textbf{Detuned RSE, no MG}}
            & \multicolumn{1}{G}{\textbf{Detuned RSE, MG}} \\
        \hline
        $\Delta g/g$       & 2\rlap{\,\%}
                           & 3\rlap{\,\%}
                           & 4\rlap{\,\%}
                           & 4\rlap{\,\%}
                           \\
        $\Delta z/z$       & \multicolumn{1}{c}{---}
                           & \multicolumn{1}{c}{---}
                           & 7\rlap{\,\%}
                           & 7\rlap{\,\%}
                           \\
        $\Delta|p|/|p|$    & 20\rlap{\,\%}
                           & 18\rlap{\,\%}
                           & 1.1\rlap{\,\%}
                           & 1.0\rlap{\,\%}
                           \\
        $\Delta Q_p / Q_p$   & \multicolumn{1}{c}{---}
                           & \multicolumn{1}{c}{---}
                           & 4\rlap{\,\%}
                           & 3\rlap{\,\%}
                           \\
        $\Delta \xi^2/\xi^2$ & \multicolumn{1}{c}{---}
                           & \multicolumn{1}{c}{---}
                           & 3\rlap{\,\%}
                           & 3\rlap{\,\%}
                           \\
        $\Delta a/a$       & 3\rlap{\,\%}
                           & 3\rlap{\,\%}
                           & 0.6\rlap{\,\%}
                           & 0.6\rlap{\,\%}
                           \\
    \end{tabular}
    \end{ruledtabular}
    \label{tab:CalRequirements}
\end{table}

\subsubsection{Massive graviton}

Here we consider parameter estimation in which the mass of the graviton (via the parameter $B$ defined earlier) is included in the parameter estimation, so that the vector of compact binary coalescenece signal parameters is $\bld{\theta} = \begin{pmatrix} \mathcal{M} & \eta & t_\text{c} & \phi_\text{c} & D & B \end{pmatrix}^\intercal$.

\begin{equation}
    \begin{bmatrix}
        \Delta \mathcal{M}/\mathcal{M} \\
        \Delta \eta /\eta \\
        \Delta t_\text{c} / (\SI{1}{\ms}) \\
        \Delta \phi_\text{c} / (\SI{1}{\rad}) \\
        \Delta D/D \\
        \Delta B / (\SI{1}{\kHz})
    \end{bmatrix}
    =
    \frac{1}{10^3}\times
    \left[
    \begin{array}{d{6.0} d{6.0} d{6.-1}}
        -29 & -34 & -84 \\
        -100 & -85 & -215 \\
        -1988 & -307 & -4066 \\
        -16317 & -19135 & -49625 \\
        720 & -103 & -621 \\
        -26 & -20 & -131 \\
    \end{array}
    \right]
    \begin{bmatrix}
        \Delta g/g \\
        \Delta |p|/|p| \\
        \Delta a/a
    \end{bmatrix}\!.
    \label{eq:CeJacZeroDetMG}
\end{equation}

The Cram\'{e}r--Rao-limited standard deviations on the astrophysical parameters are
\begin{subequations}
\begin{align}
		\Sigma_{\mathcal{M}\mathcal{M}}^{\, 1/2}/\mathcal{M} &= 0.022 \\
		\Sigma_{\eta\eta}^{\, 1/2}/\eta &= 0.051 \\
		\Sigma_{t_\text{c} t_\text{c}}^{\, 1/2} &= 0.51\,\text{ms} \\
		\Sigma_{\phi_\text{c} \phi_\text{c}}^{\, 1/2} &= 13.0\,\text{rad} \\
		\Sigma_{DD}^{\, 1/2}/D &= 0.056 \\
		\Sigma_{BB}^{\, 1/2} &= 0.024\,\text{kHz}
\end{align}
\end{subequations}

Since our parameter estimation should return $B = 0$ in the absence of error, the error $\Sigma_{BB}^{\,1/2}$ induced by the noise can be converted into an error $\Sigma_{m_G m_G}^{\,1/2}$ on the graviton mass or an error $\Sigma_{\Lambda\Lambda}^{\,1/2}$ on the graviton Compton wavelength via Eq.~\ref{eq:MgPhase}:
\begin{align}
    \Sigma_{m_G m_G}^{\,1/2} &= 2\hbar\sqrt{\frac{1+Z}{1-Z}\times\frac{\pi \Sigma_{BB}^{\,1/2}}{Dc^3}}
    \label{eq:BtoMG} \\
    \Sigma_{\Lambda\Lambda}^{\,1/2} &= \sqrt{\frac{1-Z}{1+Z}\times\frac{\pi D c}{\Sigma_{BB}^{\,1/2}}},
    \label{eq:BtoLam}
\end{align}
where we have used the result from Will~\cite{Will1998} that $\mathcal{D} \simeq (1-Z)D$ for $Z \ll 1$.
Therefore, the noise-induced error $\Sigma_{BB}^{\,1/2} = \SI{0.024}{\kHz}$ translates to an error on the graviton mass of $\Sigma_{m_G m_G}^{\,1/2} = \SI{8.0e-59}{\kg} = \SI{4.5e-23}{\eV}/c^2$, and an error on the graviton Compton wavelength of $\Sigma_{\Lambda\Lambda}^{\,1/2} = \SI{2.8e13}{\km}$~\footnote{Note that in the noiseless limit ($\rho \rightarrow \infty$), $\Sigma_{m_G m_G} \rightarrow 0$ but $\Sigma_{\Lambda\Lambda} \rightarrow \infty$.}.
These Cram\'{e}r--Rao limits are combined with the matrix in Eq.~\ref{eq:CeJacZeroDetMG} to produce the calibration limits given in Tab.~\ref{tab:CalRequirements}; these requirements are almost identical to the requirements for the massless graviton analysis, indicating that an analysis with a massive graviton does not require a more stringent calibration effort.

Eqs.~\ref{eq:BtoMG} and~\ref{eq:BtoLam} apply equally well as relations between the errors $\Delta B$, $\Delta m_G$, and $\Delta\Lambda$ introduced by calibration systematics; thus, if the calibration systematics are kept sufficiently small that $\Delta B \big/ \Sigma_{BB}^{\,1/2} < 1/3$, the corresponding error on the graviton mass is $\Delta m_G \big/ \Sigma_{m_G m_G}^{\,1/2} = \sqrt{\Delta B \big/ \Sigma_{BB}^{\,1/2}} > \sqrt{1/3} \simeq 0.6$, and on the graviton wavelength is $\Delta\Lambda / \Sigma_{\Lambda\Lambda}^{\,1/2} > \sqrt{3} \simeq 1.7$.

\subsection{Detuned signal extraction configuration, massless graviton}

This configuration employs both detuned sideband extraction and a non-\SI{90}{\degree} homodyne angle, with parameters given in Table~\ref{tab:Configurations}.
Here the vector of calibration parameters is $\bld{\lambda} = \begin{pmatrix} g & z & |p| & Q_p & \xi^2 & a \end{pmatrix}^\intercal$, and the vector of astrophysical parameters is $\bld{\theta} = \begin{pmatrix} \mathcal{M} & \eta & t_\text{c} & \phi_\text{c} & D \end{pmatrix}^\intercal$.

\begin{widetext}

The resulting matrix relation between $\Delta\bld{\hat{\theta}}$ and $\Delta\bld{\lambda}$ is
\begin{equation}
    \begin{bmatrix}
        \Delta \mathcal{M}/\mathcal{M} \\
        \Delta \eta /\eta \\
        \Delta t_\text{c} / (\SI{1}{\ms}) \\
        \Delta \phi_\text{c} / (\SI{1}{\rad}) \\
        \Delta D/D
    \end{bmatrix}
    =
    \frac{1}{10^3}\times
    \left[
    \begin{array}{d{5.1} d{5.1} d{5.1} d{5.1} d{5.1} d{5.0}}
  -16 & -2 & 15 & -15 & -11 & -21 \\
  -83 & -16 & 132 & -66 & -82 & -133 \\
  29 & -514 & 3376 & 419 & -937 & 93 \\
  -7003 & -629 & 5319 & -7329 & -3909 & -8253 \\
  134 & 73 & -512 & -114 & -194 & -1031 \\
    \end{array}
    \right]
    \begin{bmatrix}
        \Delta g/g \\
        \Delta z/z \\
        \Delta |p|/|p| \\
        \Delta Q_p/Q_p \\
        \Delta \xi^2 / \xi^2 \\
        \Delta a/a
    \end{bmatrix}.
    \label{eq:CeJacDetNoMG}
\end{equation}

The corresponding Cram\'{e}r--Rao limited estimates of the astrophysical parameters are
\begin{subequations}
\begin{align}
    \Sigma_{\mathcal{M}\mathcal{M}}^{\,1/2}/\mathcal{M} &= 0.0020 \\
    \Sigma_{\eta\eta}^{\,1/2}/\eta &= 0.0099 \\
    \Sigma_{t_\text{c} t_\text{c}}^{\,1/2} &= 0.11\,\text{ms} \\
    \Sigma_{\phi_\text{c} \phi_\text{c}}^{\,1/2} &= 0.98\,\text{rad} \\
    \Sigma_{DD}^{\,1/2}/D &= 0.017,
\end{align}
\end{subequations}
and the resulting limits on the systematic errors $\Delta\bld{\lambda}$ are again given in Tab.~\ref{tab:CalRequirements}.
Additionally, the requirements are given for the additional case of parameter estimation with a massive graviton, although there is little difference compared to the massless graviton requirements.
\end{widetext}

\section{Discussion and Conclusion}
\label{sec:Discussion}

Tab.~\ref{tab:CalRequirements} shows that in order to achieve noise-limited systematics for a GW150914-like signal detected by a single Advanced LIGO instrument running in an O1/O2 configuration, calibration accuracy of a few percent is required for the optical gain $g$ and actuation strength $a$, and about \SI{20}{\%} accuracy is required on the optical pole $|p|$.
If the detector is instead running in a detuned configuration with a higher sensitivity, there are a greater number of calibration parameters that must be characterized, and the required accuracy ranges from \SI{7}{\%} at the least stringent (for the homodyne zero $z$) to \SI{0.6}{\%} at the most stringent (for the actuation strength).
The inclusion of a model with nonzero graviton mass does not significantly alter these requirements.

There are several obvious extensions of this work.
The results presented here rely on a single detector, with a waveform model that does not include the spins, location, and orientation of the system.
The effects of these additional parameters should be examined, using a multiple-detector configuration.
The analysis should be repeated on a wide variety of coalescence systems (different component masses, spins, distances, etc.) in order to set calibration requirements that are sufficient for the majority of the coalescences that are expected to be detected with Advanced LIGO.

This semianalytical approach could be complemented by a fully numerical analysis in which one examines the effect of systematic errors in the calibration parameters on the full Advanced LIGO parameter estimation pipeline~\cite{GW150914PE}.
This would allow requirements on the calibration parameters to be set even for signals with SNR $\rho \lesssim 20$, where the Cram\'{e}r--Rao analysis is no longer valid.

Finally, this analysis could be useful in determining calibration requirements for future generations of gravitational-wave detectors~\cite{Punturo2010,Evans2017}.
These detectors are expected to be more sensitive by a factor of 10 or more in amplitude compared to Advanced LIGO, requiring similar improvements in their calibration accuracy.

\begin{acknowledgments}
LIGO was constructed by the California Institute of Technology and Massachusetts Institute of Technology with funding from the National Science Foundation and operates under cooperative agreement PHY--0757058.
This work has internal LIGO document number P1700033.
\end{acknowledgments}

\bibliographystyle{aipnum4-1}
\bibliography{CalSysErrPE.bib}
\end{document}